# A Spatio-temporal Decomposition Method for the Coordinated Economic Dispatch of Integrated Transmission and Distribution Grids

Qi Wang, Wenchuan Wu, *Fellow, IEEE*, Chenhui Lin, Bin Wang

*Abstract*—With numerous distributed energy resources (DERs) integrated into the distribution networks (DNs), the coordinated economic dispatch (C-ED) is essential for the integrated transmission and distribution grids. For large scale power grids, the centralized C-ED meets high computational burden and information privacy issues. To tackle these issues, this paper proposes a spatio-temporal decomposition algorithm to solve the C-ED in a distributed and parallel manner. In the temporal dimension, the multi-period economic dispatch (ED) of transmission grid (TG) is decomposed to several subproblems by introducing auxiliary variables and overlapping time intervals to deal with the temporal coupling constraints. Besides, an accelerated alternative direction method of multipliers (A-ADMM) based temporal decomposition algorithm with the warm-start strategy, is developed to solve the ED subproblems of TG in parallel. In the spatial dimension, a multi-parametric programming projection based spatial decomposition algorithm is developed to coordinate the ED problems of TG and DNs in a distributed manner. To further improve the convergence performance of the spatial decomposition algorithm, the aggregate equivalence approach is used for determining the feasible range of boundary variables of TG and DNs. Moreover, we prove that the proposed spatio-temporal decomposition method can obtain the optimal solution for bilevel convex optimization problems with continuously differentiable objectives and constraints. Numerical tests are conducted on three systems with different scales, demonstrating the high computational efficiency and scalability of the proposed spatio-temporal decomposition method.

## Nomenclature

### A. Superscripts

| | |
|---|---|
| $(\cdot)^{\text{trans}}$ | Sets/variables/parameters of the TG |
| $(\cdot)^{\text{dist},k}$ | Sets/variables/parameters of the $k^{\text{th}}$ DN |

### B. Sets

| | |
|---|---|
| $T$ | Index set of the whole optimization horizon |
| $G^{\text{trans}}$ | Index set of generator bus numbers of the TG |
| $DIST$ | Index set of all DNs |
| $DG, E$ | Index sets of distributed generation (DG) and energy storage system (ESS) bus numbers |
| $B^{\text{trans}}$ | Index set of boundary bus numbers of the TG |
| $D$ | Index set of load bus numbers |
| $L$ | Index set of line numbers |
| $N^{\text{dist},k}$ | Index set of bus numbers of the $k^{\text{th}}$ DN |
| $I_{\text{TL}}^{\text{dist},k}, I_{\text{TL},j}^{\text{dist},k}$ | Index set of thermal load (TL) bus numbers and the TLs connected to bus $j$ of the $k^{\text{th}}$ DN |

### C. Parameters

| | |
|---|---|
| $T_{\text{hor}}$ | The whole optimization horizon |
| $\Delta T$ | Optimization time interval |
| $\overline{P}_{\text{DG},i}, \underline{P}_{\text{DG},i}$ | Maximum/minimum active power output of DG at bus $i$ |
| $\eta_{\text{ESS},i}^{\text{ch}}, \eta_{\text{ESS},i}^{\text{dc}}$ | Charging and discharging efficiency of the ESS at bus $i$ |
| $P_{\text{D},i}(t)$ | Active power of uncontrollable load at bus $i$ during period $t$ |
| $\overline{P}_{\text{L},j}^{\text{trans}}$ | Line flow limit of line $j$ of the TG |
| $SF_{j \to i}^{\text{trans}}$ | Shift factor for bus $i$ on line $j$ of the TG |
| $RU_i^{\text{trans}}, RD_i^{\text{trans}}$ | Upward/downward ramp rate of generator at bus $i$ of the TG |
| $\overline{P}_{\text{G},i}^{\text{trans}}, \underline{P}_{\text{G},i}^{\text{trans}}$ | Maximum/minimum active power output of generator at bus $i$ of the TG |
| $SRU^{\text{trans}}(t), SRD^{\text{trans}}(t)$ | System-wide upward/downward spinning reserve capacity requirement during period $t$ of the TG |
| $E_{\text{ESS},i}(0), \underline{E}_{\text{ESS},i}, \overline{E}_{\text{ESS},i}$ | Initial capacity, lower and upper limits of the capacity of the ESS connected to bus $i$ |
| $\overline{P}_{\text{ESS},i}^{\text{ch}}, \overline{P}_{\text{ESS},i}^{\text{dc}}$ | Upper limits of charging and discharging active power of the ESS connected to bus $i$ |
| $\underline{p}_{\text{B},i}^{\text{trans}}(t), \overline{p}_{\text{B},i}^{\text{trans}}(t)$ | Lower and upper limits of the boundary active power at bus $i$ of the TG during period $t$ |
| $\hat{P}_{i \to j}^{\text{dist},k}(t), \hat{Q}_{i \to j}^{\text{dist},k}(t), \hat{V}_i^{\text{dist},k}(t)$ | Operational base points of active/reactive power from bus $i$ to bus $j$ and voltage magnitude of bus $i$ during period $t$ of the $k^{\text{th}}$ DN |
| $R_{i \to j}^{\text{dist},k}$ | Line resistance from bus $i$ to $j$ of the $k^{\text{th}}$ DN |
| $\overline{P}_{\text{L},i \to j}^{\text{dist},k}$ | Line flow limit of line from bus $i$ to bus $j$ of the $k^{\text{th}}$ DN |
| $T_{\text{out}}^{\text{dist},k}(t)$ | Outdoor temperature at period $t$ of the $k^{\text{th}}$ DN |
| $\underline{T}_{\text{in},h}^{\text{dist},k}, \overline{T}_{\text{in},h}^{\text{dist},k}$ | Lower and upper limits of the indoor temperature in the $h^{\text{th}}$ household of the $k^{\text{th}}$ DN |

### D. Variables

| | |
|---|---|
| $p_{\text{G},i}^{\text{trans}}(t)$ | Active power output of generator at bus $i$ during period $t$ of the TG |

Manuscript received XX, 2022. This work was supported by the National Science Foundation of China under Grant 51725703. (Corresponding Author: Wenchuan Wu).
Q. Wang, W. Wu (email: wuwench@tsinghua.edu.cn), C. Lin and B. Wang are with the Department of Electrical Engineering, Tsinghua University, Beijing 100084, China.

| Symbol | Description |
|---|---|
| $p_{\text{DG},i}(t)$ | Active power output of DG at bus $i$ during period $t$ |
| $P_{\text{ESS},i}^{\text{dc}}(t)$, $P_{\text{ESS},i}^{\text{ch}}(t)$ | Discharging and charging active power of the ESS connected to bus $i$ during period $t$ |
| $p_{\text{B},i}^{\text{trans}}(t)$, $p_{\text{B}}^{\text{dist},k}(t)$ | Transferred active power at bus $i$ of the TG to DNs and that received by the $k^{\text{th}}$ DN from TG during period $t$ |
| $ru_i^{\text{trans}}(t)$, $rd_i^{\text{trans}}(t)$ | Upward/downward spinning reserve contributions of generator at bus $i$ during period $t$ of the TG |
| $E_{\text{ESS},i}(t)$ | The state-of-charge (SOC) of the ESS connected to bus $i$ during period $t$ |
| $p_{i \to j}^{\text{dist},k}(t)$, $l_{i \to j}^{\text{dist},k}(t)$ | Line flow and loss from bus $i$ to bus $j$ during period $t$ of the $k^{\text{th}}$ DN |
| $p_j^{\text{dist},k}(t)$ | Net injected active power at bus $j$ during period $t$ of the $k^{\text{th}}$ DN |
| $p_{\text{TL},h}^{\text{dist},k}(t)$ | Active power of the thermal load in the $h^{\text{th}}$ household during period $t$ of the $k^{\text{th}}$ DN |
| $T_{\text{in},h}^{\text{dist},k}(t)$ | Indoor temperature in the $h^{\text{th}}$ household during period $t$ of the $k^{\text{th}}$ DN |

## I. INTRODUCTION

### A. Background

TO fully exploit the flexibility on the DNs side and accommodate the integration of DERs, the coordinated operation between the transmission and distribution networks is indispensable. The multi-period C-ED problem of the integrated transmission and distribution grids (hereafter named ***C-ED problem***), is a dynamic optimization model essentially, which is used for the day-ahead and the intraday look-ahead power dispatch. With the increasing integration of DERs into the transmission and distribution grids, some key issues are raised:

i) Since the high computational burden brought by the large-scale power grids and numerous DERs, the C-ED needs the algorithms with the properties of fast convergence and scalability.

ii) With the increasing penetration of renewable DGs, the transmission congestion problem becomes more serious. Correspondingly, this will make more operational constraints active in the ED optimization problems, resulting in a substantial growth in the computational effort.

iii) As more and more ESSs and thermal loads are integrated into power systems, the power grid operation presents strong temporal coupling characteristics, which increases the computational burden and deteriorates the convergence performance of the conventional distributed algorithms.

The existing distributed algorithms for solving the C-ED problem of the integrated transmission and distribution grids are all the spatial decomposition algorithms, and generally only have first-order convergence. This makes it quite challenging for the traditionally spatially distributed algorithms to fully meet the actual efficiency requirements of the aforementioned issues. To satisfy the operation requirements, this work concentrates on developing a coordination algorithm combining the spatial decomposition and temporal decomposition to enhance the convergence performance.

### B. Previous Research

The coordinated operation of the integrated transmission and distribution grids has been studied by some scholars recently. The distributed optimization methods that are utilized to solve the C-ED problem, are summarized here, which can be roughly divided into three categories: i) the primal decomposition algorithms, represented by the Benders decomposition (GBD) method [1]-[2]; ii) the dual decomposition algorithms, including the alternating direction method of multipliers (ADMM) [3]-[4], analytical target cascading (ATC) [5]-[6], and augmented Lagrangian (ALR) [7]; iii) the Karush-Kuhn-Tucker (KKT) condition based methods, such as heterogeneous decomposition (HGD) [9]-[10].

For the primal decomposition algorithms, a hierarchical coordination scheme was proposed in [1] for the C-ED adopting the GBD algorithm. Reference [2] developed a modified GBD to decompose the C-ED problem. However, their efficiency is relatively low, since every DN only generates one feasible cut per iteration for infeasible cases. Moreover, the optimization problems of DNs are temporal coupling within the whole scheduling horizon, which leads to heavy computational burden for generating the feasible cuts.

When it comes to the dual decomposition algorithms, a coordinated robust dynamic C-ED model was proposed in [3] solving by ADMM. Reference [4] developed an ADMM-based robust decentralized coordination schedule method, which achieved a balance between the operational costs and risk for the integrated transmission and distribution grids. An ATC-based distributed scheduling solution was developed in [5] to achieve real-time coordinated schedule for the integrated transmission and distribution grids, in which the uncertainties of renewable DGs were modeled as intervals. A novel dynamic economic dispatch model was proposed in [6], which could make a tradeoff between economic cost and reliability cost utilizing ATC. In [7], the C-ED problem was formulated as a two-stage robust model to capture the uncertainties of renewable DGs and solved by an ALR based method. It is worth to note that the dual decomposition methods have the hyperparameter tuning issues, which significantly affects the convergence performance [8]. Moreover, the dual decomposition algorithms, such as ADMM and ATC, have only first-order convergence.

As for the KKT condition based methods, references [9] and [10] proposed a HGD method to solve the C-ED problems, wherein the locational marginal price at the boundary bus of TG and DNs' boundary power were interacted between TG and DNs. Though the HGD usually shows good convergence, its convergence region is relatively small [11].

What's more, it is noteworthy that the computational burden of multi-period ED increases dramatically with the growing scales of the power systems, the number of DERs, and the temporal coupling constraints [12]. An intuitive idea is to decompose large problems into smaller subproblems adopting distributed algorithms to accelerate the computation procedure.

Along this line of thought, there are some related researches on the temporal decomposition strategy. A stochastic model predictive control method was proposed in [13] to schedule the dispatchable generation, in which a comprehensive strategy of temporary decomposition and scenario decomposition, was introduced to reduce the solution time. A temporal decomposition strategy to decrease the computational time of security-constrained unit commitment was presented in [14]. However, the aforementioned literature all focuses on the ED of power systems at the same level. To the best of our knowledge, the temporal decomposition strategy has not been investigated in the bilevel C-ED problem of the integrated transmission and distribution grids.

Moreover, some crucial issues are raised when the C-ED problem is implemented in a temporal decomposition manner with the increasing integration of numerous ESSs and thermal loads into power systems:
  i) How to deal with the temporal coupling constraints in TG, including the generators' ramp up and down constraints and the ESS energy constraints, without changing the optimal solution?
  ii) How to modify the decentralized algorithm to ensure the optimality of the obtained solution for the C-ED problem after implementing the temporal decomposition?
  iii) Considering the strong temporal coupling characteristics of power systems, the convergence performance of the existing algorithms is deteriorated. Whether more efficient algorithms are needed to enhance the convergence performance?

*C. Contributions*

To fill the aforementioned research gaps, this paper firstly describes a comprehensive C-ED model containing renewable DGs, ESSs and thermal loads for the integrated transmission and distribution grids. It is worth emphasizing that the temporal coupling flexible resources, i.e., ESSs and thermal loads, were not considered in the previous literature [1]-[7], [9]-[10].

Subsequently, we propose a novel spatio-temporal decomposition method to make the C-ED applicable for large scale transmission and distribution grids. To the best of our knowledge, such an effort has not been implemented in the previous studies. The unique contributions of the proposed method include:
  1) In the temporal dimension, the multi-period ED of TG is decomposed to several subproblems by introducing auxiliary variables and overlapping time intervals to decouple the temporal coupling constraints, including the generators' ramp up and down constraints and the ESS energy constraints. In this way, the TG's ED can be solved in parallel. Besides, a scheme is proposed to determine a proper number of subproblems and their time periods to guarantee the efficiency. Moreover, an A-ADMM based temporal decomposition algorithm adopting the Nesterov accelerated gradient descent method [15] with the warm-start strategy, is adopted to parallelly coordinate the subproblems of TG.
  2) In the spatial dimension, we modify the multi-parametric programming projection decomposition algorithm developed in our previous work [16] to adapt to the temporal decomposition framework. Specifically, the projection functions generated by the DNs are also decomposed into several sub-horizons accordingly. It should be noted that the projection functions are calculated with second-order exactness, thus the proposed multi-parametric programming projection based spatial decomposition algorithm has a superlinear convergence.
  3) To further improve the convergence performance of the spatial decomposition algorithm, the aggregate equivalence approach is used for determining the feasible range of boundary variables of TG and DNs. This strategy can make the proposed method quickly step out of the infeasible region of the C-ED problem, and speed up the iterative process. At the same time, the time of finding a feasible solution for the C-ED problem in the initial stage is significantly reduced.

We also prove that the proposed spatio-temporal decomposition method can obtain the optimal solution for bilevel convex optimization problems with continuously differentiable objectives and constraints.

The remainder is organized as follows. We introduce the mathematical model of the C-ED problem in Section II. In Section III, we not only present the temporal decomposition process of TG's multi-period ED model, but also develop the spatio-temporal decomposition algorithm, and give the convergence proof. Section IV presents the case studies and comparisons. The conclusions are drawn in Section V.

II. MATHEMATICAL MODEL FORMULATION

*A. Objective Function*

The objective for the C-ED problem is given as follows:

$$\min \left\{ \sum_{t \in T} \left( \sum_{i \in G^{\text{trans}}} C_i^{\text{trans}} \left( p_{\text{G},i}^{\text{trans}}(t) \right) + \sum_{i \in DG^{\text{trans}}} C_i^{\text{trans}} \left( p_{\text{DG},i}^{\text{trans}}(t) \right) + \sum_{i \in E^{\text{trans}}} C_i^{\text{trans}} \left( P_{\text{ESS},i}^{\text{dc,trans}}(t), P_{\text{ESS},i}^{\text{ch,trans}}(t) \right) \right) + \sum_{t \in T} \sum_{k \in DIST} \left( C_{\text{fee}}^{\text{dist},k}(t) + \sum_{j \in DG^{\text{dist},k}} C_j^{\text{dist},k} \left( p_{\text{DG},j}^{\text{dist},k}(t) \right) + \sum_{j \in E^{\text{dist},k}} C_j^{\text{dist},k} \left( P_{\text{ESS},j}^{\text{dc,dist},k}(t), P_{\text{ESS},j}^{\text{ch,dist},k}(t) \right) \right) \right\} \quad (1)$$

where $C_i^{\text{trans}} \left( p_{\text{G},i}^{\text{trans}}(t) \right)$ is the generation cost of the generator at bus $i$ of the TG. $C_i^{\text{trans}} \left( p_{\text{DG},i}^{\text{trans}}(t) \right)$ and $C_i^{\text{trans}} \left( P_{\text{ESS},i}^{\text{dc,trans}}(t), P_{\text{ESS},i}^{\text{ch,trans}}(t) \right)$ are the penalization of renewable DG's curtailment, and the penalization of simultaneous charging and discharging of ESS at bus $i$ of TG, respectively. $C_{\text{fee}}^{\text{dist},k}(t)$ is the electricity fee of the $k^{\text{th}}$ DN for purchasing electricity from TG. $C_j^{\text{dist},k} \left( p_{\text{DG},j}^{\text{dist},k}(t) \right)$ and $C_j^{\text{dist},k} \left( P_{\text{ESS},j}^{\text{dc,dist},k}(t), P_{\text{ESS},j}^{\text{ch,dist},k}(t) \right)$ are the penalization of renewable DG's output curtailment, and the penalization of simultaneous charging and discharging of ESS at bus $j$ of the $k^{\text{th}}$ DN, respectively. The detailed expressions of the above functions in formula (1) are given as follows:

$$C_i^{\text{trans}} \left( p_{\text{G},i}^{\text{trans}}(t) \right) = a_{2,i} \left( p_{\text{G},i}^{\text{trans}}(t) \right)^2 + a_{1,i} p_{\text{G},i}^{\text{trans}}(t) + a_{0,i} \quad (2)$$

where $a_{2,i}$, $a_{1,i}$ and $a_{0,i}$ are the coefficients of the quadratic, linear and constant terms, respectively.

$$C_{\text{fee}}^{\text{dist},k}(t) = b^{\text{dist}}(t) \cdot p_{\text{B}}^{\text{dist},k}(t) \tag{3}$$

where $b^{\text{dist}}(t)$ is the electricity price during period $t$.

$$C_i^{\text{trans}}\left(p_{\text{DG},i}^{\text{trans}}(t)\right) = \frac{\sigma_{\text{DG}}\left(p_{\text{DG},i}^{\text{trans}}(t) - \overline{P}_{\text{DG},i}^{\text{trans}}\right)^2}{\overline{P}_{\text{DG},i}^{\text{trans}}} \tag{4}$$

$$C_j^{\text{dist},k}\left(p_{\text{DG},j}^{\text{dist},k}(t)\right) = \frac{\sigma_{\text{DG}}\left(p_{\text{DG},j}^{\text{dist},k}(t) - \overline{P}_{\text{DG},j}^{\text{dist},k}\right)^2}{\overline{P}_{\text{DG},j}^{\text{dist},k}} \tag{5}$$

where $\sigma_{\text{DG}}$ is the penalty coefficient for renewable DGs' output curtailments.

To avoid the simultaneous charging and discharging paradox, we introduce the following penalty term into the objective:

$$C_i^{\text{trans}}\left(P_{\text{ESS},i}^{\text{dc,trans}}(t), P_{\text{ESS},i}^{\text{ch,trans}}(t)\right) = \sigma_{\text{ESS}}\left(P_{\text{ESS},i}^{\text{dc,trans}}(t)\left(\frac{1}{\eta_{\text{ESS},i}^{\text{dc,trans}}} - 1\right) + P_{\text{ESS},i}^{\text{ch,trans}}(t)\left(1 - \eta_{\text{ESS},i}^{\text{ch,trans}}\right)\right) \tag{6}$$

$$C_j^{\text{dist},k}\left(P_{\text{ESS},j}^{\text{dc,dist},k}(t), P_{\text{ESS},j}^{\text{ch,dist},k}(t)\right) = \sigma_{\text{ESS}}\left(P_{\text{ESS},j}^{\text{dc,dist},k}(t)\left(\frac{1}{\eta_{\text{ESS},j}^{\text{dc,dist},k}} - 1\right) + P_{\text{ESS},j}^{\text{ch,dist},k}(t)\left(1 - \eta_{\text{ESS},j}^{\text{ch,dist},k}\right)\right) \tag{7}$$

where $\sigma_{\text{ESS}}$ is the penalty coefficient for simultaneous charging and discharging of ESSs.

### B. Operational Constraints of the Transmission Network

The following constraints should be satisfied in the TG.

*1) Power Balance Constraints*

The sum of the outputs of generators, DGs and ESSs should be equivalent to that of the loads plus the power sent to DNs:

$$\sum_{i \in G^{\text{trans}}} p_{\text{G},i}^{\text{trans}}(t) + \sum_{i \in DG^{\text{trans}}} p_{\text{DG},i}^{\text{trans}}(t) + \sum_{i \in E^{\text{trans}}} \left(p_{\text{ESS},i}^{\text{dc,trans}}(t) - p_{\text{ESS},i}^{\text{ch,trans}}(t)\right) = \sum_{i \in B^{\text{trans}}} p_{\text{B},i}^{\text{trans}}(t) + \sum_{i \in D^{\text{trans}}} P_{\text{D},i}^{\text{trans}}(t), \forall t \in T \tag{8}$$

*2) Network Constraints*

The power on every branch should be within its capacity.

$$-\overline{P}_{\text{L},j}^{\text{trans}} \le \sum_{i \in G^{\text{trans}}} SF_{j-i}^{\text{trans}} p_{\text{G},i}^{\text{trans}}(t) + \sum_{i \in DG^{\text{trans}}} SF_{j-i}^{\text{trans}} p_{\text{DG},i}^{\text{trans}}(t) + \sum_{i \in E^{\text{trans}}} SF_{j-i}^{\text{trans}} \left(p_{\text{ESS},i}^{\text{dc,trans}}(t) - p_{\text{ESS},i}^{\text{ch,trans}}(t)\right) - \sum_{i \in B^{\text{trans}}} SF_{j-i}^{\text{trans}} p_{\text{B},i}^{\text{trans}}(t) - \sum_{i \in D^{\text{trans}}} SF_{j-i}^{\text{trans}} P_{\text{D},i}^{\text{trans}}(t) \le \overline{P}_{\text{L},j}^{\text{trans}}, \forall j \in L^{\text{trans}}, \forall t \in T \tag{9}$$

*3) Spinning Reserve Constraints*

Sufficient spinning reserve is required for TG.

$$0 \le ru_i^{\text{trans}}(t) \le RU_i^{\text{trans}} \Delta T,$$
$$ru_i^{\text{trans}}(t) \le \overline{P}_{\text{G},i}^{\text{trans}} - p_{\text{G},i}^{\text{trans}}(t), \forall i \in G^{\text{trans}}, \forall t \in T \tag{10}$$

$$0 \le rd_i^{\text{trans}}(t) \le RD_i^{\text{trans}} \Delta T,$$
$$rd_i^{\text{trans}}(t) \le p_{\text{G},i}^{\text{trans}}(t) - \underline{P}_{\text{G},i}^{\text{trans}}, \forall i \in G^{\text{trans}}, \forall t \in T \tag{11}$$

$$\sum_{i \in G^{\text{trans}}} ru_i^{\text{trans}}(t) \ge SRU^{\text{trans}}(t), \sum_{i \in G^{\text{trans}}} rd_i^{\text{trans}}(t) \ge SRD^{\text{trans}}(t), \forall t \in T \tag{12}$$

*4) Ramping Constraints*

Variations of generators' outputs between adjacent dispatch periods should be within the generators' ramping limits.

$$-RD_i^{\text{trans}} \Delta T \le p_{\text{G},i}^{\text{trans}}(t) - p_{\text{G},i}^{\text{trans}}(t-1) \le RU_i^{\text{trans}} \Delta T, \forall t \in T \setminus \{1\} \tag{13}$$

*5) Generator's Output Constraints*

The generators' outputs should be kept within the bounds.

$$\underline{P}_{\text{G},i}^{\text{trans}} \le p_{\text{G},i}^{\text{trans}}(t) \le \overline{P}_{\text{G},i}^{\text{trans}}, \forall i \in G^{\text{trans}}, \forall t \in T \tag{14}$$

*6) DG's Output Constraints*

The active power of renewable DGs should also be within the lower and upper bounds.

$$\underline{P}_{\text{DG},i}^{\text{trans}} \le p_{\text{DG},i}^{\text{trans}}(t) \le \overline{P}_{\text{DG},i}^{\text{trans}}, \forall i \in DG^{\text{trans}}, \forall t \in T \tag{15}$$

*7) ESS's Operational Constraints*

The ESSs should satisfy the following operational constraints.

$$E_{\text{ESS},i}^{\text{trans}}(t) = E_{\text{ESS},i}^{\text{trans}}(t-1) + \left(p_{\text{ESS},i}^{\text{ch,trans}}(t) \times \eta_{\text{ESS},i}^{\text{ch,trans}} - p_{\text{ESS},i}^{\text{dc,trans}}(t) / \eta_{\text{ESS},i}^{\text{dc,trans}}\right) \Delta T, \forall i \in E^{\text{trans}}, \forall \tau \in T \tag{16}$$

$$\underline{E}_{\text{ESS},i}^{\text{trans}} \le E_{\text{ESS},i}^{\text{trans}}(t) \le \overline{E}_{\text{ESS},i}^{\text{trans}}, \forall j \in E^{\text{trans}}, \forall \tau \in T \tag{17}$$

$$E_{\text{ESS},i}^{\text{trans}}(T_{\text{hor}}) = E_{\text{ESS},i}^{\text{trans}}(0), \forall j \in E^{\text{trans}} \tag{18}$$

$$0 \le p_{\text{ESS},i}^{\text{ch,trans}}(t) \le \overline{P}_{\text{ESS},i}^{\text{ch,trans}}, 0 \le p_{\text{ESS},i}^{\text{dc,trans}}(t) \le \overline{P}_{\text{ESS},i}^{\text{dc,trans}} \tag{19}$$

$$p_{\text{ESS},i}^{\text{ch,trans}}(t) \times p_{\text{ESS},i}^{\text{dc,trans}}(t) = 0, \forall j \in E^{\text{trans}}, \forall t \in T \tag{20}$$

To prevent the simultaneous charging and discharging paradox, the complementary constraint (20) is introduced. While notice that (20) essentially is a bilinear equality which is hard to solve. To deal with this dilemma, (20) can be omitted by adding a penalty term (6) to the objective function. Refer to [17] for detailed explanations.

*8) Limits on the Boundary Active Power*

Generally, a boundary power range can be roughly estimated. In order to quickly jump out of the infeasible region of the C-ED problem and speed up the iterative process, we utilize the aggregate equivalence approach to determine $\underline{p}_{\text{B},i}^{\text{trans}}(t)$ and $\overline{p}_{\text{B},i}^{\text{trans}}(t)$. The detailed explanations are given in subsection D of Section III.

$$\underline{p}_{\text{B},i}^{\text{trans}}(t) \le p_{\text{B},i}^{\text{trans}}(t) \le \overline{p}_{\text{B},i}^{\text{trans}}(t), \forall i \in B^{\text{trans}}, \forall t \in T \tag{21}$$

### C. Operational Constraints of Distribution Network

Considering that the DC power flow model may bring significant errors in DNs, a linearized AC power flow model is applied in the DNs' models. Please refer to [18] for further details of the linearized AC power flow model.

*1) Power Flow Constraints*

$$\sum_{i:i \to j} \left(p_{i \to j}^{\text{dist},k}(t) - l_{i \to j}^{\text{dist},k}(t)\right) + p_j^{\text{dist},k}(t) = \sum_{m:j \to m} p_{j \to m}^{\text{dist},k}(t), \forall j \in N^{\text{dist},k} \tag{22}$$

$$p_j^{\text{dist},k}(t) = \begin{cases} p_{\text{DG},j}^{\text{dist},k}(t) + \left(p_{\text{ESS},j}^{\text{dc,dist},k}(t) - p_{\text{ESS},j}^{\text{ch,dist},k}(t)\right) - \sum_{h \in I_{\text{TL},j}^{\text{dist},k}} p_{\text{TL},h}^{\text{dist},k}(t) \\ \quad - P_{\text{D},j}^{\text{dist},k}(t) + p_{\text{B}}^{\text{dist},k}(t), j \text{ is root bus} \\ p_{\text{DG},j}^{\text{dist},k}(t) + \left(p_{\text{ESS},j}^{\text{dc,dist},k}(t) - p_{\text{ESS},j}^{\text{ch,dist},k}(t)\right) - \sum_{h \in I_{\text{TL},j}^{\text{dist},k}} p_{\text{TL},h}^{\text{dist},k}(t) \\ \quad - P_{\text{D},j}^{\text{dist},k}(t), j \text{ is not root bus} \end{cases} \tag{23}$$

$$l_{i \to j}^{\text{dist},k}(t) = \left[\left(\hat{P}_{i \to j}^{\text{dist},k}(t)\right)^2 + \left(\hat{Q}_{i \to j}^{\text{dist},k}(t)\right)^2\right] R_{i \to j}^{\text{dist},k} / \left(\hat{V}_i^{\text{dist},k}(t)\right)^2 + 2\left(p_{i \to j}^{\text{dist},k}(t) - \hat{P}_{i \to j}^{\text{dist},k}(t)\right) \hat{P}_{i \to j}^{\text{dist},k}(t) R_{i \to j}^{\text{dist},k} / \left(\hat{V}_i^{\text{dist},k}(t)\right)^2 \tag{24}$$

*2) Line Transmission Capacity Constraints*

$$-\overline{P}_{\text{L},i \to j}^{\text{dist},k} \le p_{i \to j}^{\text{dist},k}(t) \le \overline{P}_{\text{L},i \to j}^{\text{dist},k}, \forall (i \to j) \in L^{\text{dist},k}, \forall t \in T \tag{25}$$

*3) DG's Output Constraints*

$$\underline{P}_{\text{DG},i}^{\text{dist},k} \le p_{\text{DG},i}^{\text{dist},k}(t) \le \overline{P}_{\text{DG},i}^{\text{dist},k}, \forall i \in DG^{\text{dist},k}, \forall t \in T \tag{26}$$

### 4) ESS's Operational Constraints

$$E_{\text{ESS},j}^{\text{dist},k}(t) = E_{\text{ESS},j}^{\text{dist},k}(t-1) + \left(p_{\text{ESS},j}^{\text{ch,dist},k}(t) \times \eta_{\text{ESS},j}^{\text{ch,dist},k} - p_{\text{ESS},j}^{\text{dc,dist},k}(t)/\eta_{\text{ESS},j}^{\text{dc,dist},k}\right)\Delta T, \forall j \in E^{\text{dist},k}, \forall \tau \in T \quad (27)$$

$$\underline{E}_{\text{ESS},j}^{\text{dist},k} \leq E_{\text{ESS},j}^{\text{dist},k}(t) \leq \overline{E}_{\text{ESS},j}^{\text{dist},k}, \forall j \in E^{\text{dist},k}, \forall \tau \in T \quad (28)$$

$$E_{\text{ESS},j}^{\text{dist},k}(T_{\text{hor}}) = E_{\text{ESS},j}^{\text{dist},k}(0), \forall j \in E^{\text{dist},k} \quad (29)$$

$$0 \leq p_{\text{ESS},j}^{\text{ch,dist},k}(t) \leq \overline{P}_{\text{ESS},j}^{\text{ch,dist},k}, 0 \leq p_{\text{ESS},j}^{\text{dc,dist},k}(t) \leq \overline{P}_{\text{ESS},j}^{\text{dc,dist},k} \quad (30)$$

$$p_{\text{ESS},j}^{\text{ch,dist},k}(t) \times p_{\text{ESS},j}^{\text{dc,dist},k}(t) = 0, \forall j \in E^{\text{dist},k}, \forall t \in T \quad (31)$$

Same with equation (20), (31) can be omitted.

### 5) Thermal Load's Operational Constraints

Thermal loads should meet the thermodynamic equation [19], and ensure the indoor temperature is within the comfort range.

$$T_{\text{in},h}^{\text{dist},k}(t) = T_{\text{in},h}^{\text{dist},k}(t-1) + m_h^{\text{dist},k}\left(T_{\text{out}}^{\text{dist},k}(t) - T_{\text{in},h}^{\text{dist},k}(t-1)\right) + l_h^{\text{dist},k} p_{\text{TL},h}^{\text{dist},k}(t), \forall h \in I_{\text{TL}}^{\text{dist},k}, \forall t \in T \setminus \{1\} \quad (32)$$

$$\underline{T}_{\text{in},h}^{\text{dist},k} \leq T_{\text{in},h}^{\text{dist},k}(t) \leq \overline{T}_{\text{in},h}^{\text{dist},k} \quad (33)$$

where $m_h^{\text{dist},k}$ and $l_h^{\text{dist},k}$ are constant parameters. And $l_h^{\text{dist},k}$ is the parameter used to control heating or cooling for thermal loads.

### D. Boundary Constraints

The transferred active power between the TG and the DNs during every scheduling period should be consistent.

$$p_{\text{B},I(k)}^{\text{trans}}(t) = p_{\text{B}}^{\text{dist},k}(t), \forall k \in DIST, \forall t \in T \quad (34)$$

where $I(k)$ denotes the TG's bus index connected by the $k^{\text{th}}$ DN.

### E. Reformulation of the C-ED Problem of the Integrated Transmission and Distribution Grids

The C-ED model is a quadratic convex optimization problem with linear constraints. The variables related to TG can further be classified into internal variables $x^{\text{trans}}$ and boundary variables $l^{\text{trans}}$. $l^{\text{trans},k}$ denotes the sub-vector of $l^{\text{trans}}$ related to the $k^{\text{th}}$ DN. The variables of the $k^{\text{th}}$ DN includes internal variables $x^{\text{dist},k}$ and boundary variables $u^{\text{dist},k}$. In this way, the C-ED model can be rewritten as follows:

$$(\text{P1}) \quad \min C^{\text{trans}}\left(x^{\text{trans}}\right) + \sum_{k \in DIST} C^{\text{dist},k}\left(x^{\text{dist},k}\right) \quad (35)$$

$$s.t. \quad G^{\text{trans}}\left(x^{\text{trans}}, l^{\text{trans}}\right) \leq \mathbf{0} \quad (36)$$

$$G^{\text{dist},k}\left(x^{\text{dist},k}, u^{\text{dist},k}\right) \leq \mathbf{0}, \forall k \in DIST \quad (37)$$

$$u^{\text{dist},k} = l^{\text{trans},k}, \forall k \in DIST \quad \left(\lambda_{\text{Bou}}^{\text{dist},k}\right) \quad (38)$$

where $C^{trans}(\cdot)$ and $C^{dist,k}(\cdot)$ are the objective functions of the TG and the $k^{\text{th}}$ DN, respectively. Constraint (36) represents the feasible region of TG, including formulas (8)-(19), (21). Constraint (37) corresponds to equations (22)-(30), (32)-(33) of the $k^{\text{th}}$ DN. Constraint (38) refers to the boundary constraint (34). $\mathbf{0}$ is the null vector. $\lambda_{\text{Bou}}^{\text{dist},k}$ is the dual multiplier.

## III. SPATIO-TEMPORAL DECOMPOSITION ALGORITHM

### A. The Spatio-temporal Decomposition Framework for the C-ED Problem

Fig. 1 presents the proposed spatio-temporal decomposition framework for the C-ED problem, wherein $SP_n$ denotes the $n^{\text{th}}$ subproblem. To be specific, the spatial decomposition is adopted to solve the bilevel optimization problems between TG and DNs. And the temporal decomposition is mainly used in solving the multi-period ED of TG for its large scale. See the subsequent subsections for detailed explanations.

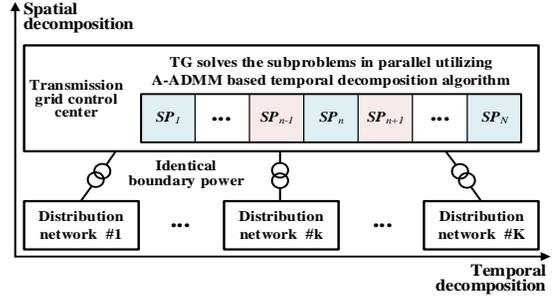

Fig. 1. Illustrating the spatio-temporal decomposition framework for the C-ED problem of the integrated transmission and distribution grids.

### B. The Spatial Decomposition Algorithm for the C-ED Problem

To coordinate the TG and DNs, we introduce the multi-parametric programming projection decomposition algorithm proposed in our previous research [16]. During the coordination process, the TG first transmits the values of the boundary variables to the connected DNs. Subsequently, every DN solves its local optimization problem under the given boundary variable values and returns a first-order optimal cut and a second-order projection function to the TG. Then, the TG coordinates the first-order optimal cuts and second-order projection functions collected from DNs, and transmits the latest boundary variables to DNs. The algorithm iterates between TG and DNs until the convergence criterion is satisfied.

While it is worthwhile to highlight that there are still some issues need to be addressed to embed this algorithm into the proposed temporal decomposition framework:

i) The first-order optimal cuts and second-order projection functions generated by DNs need to be decomposed into several sub-horizons to adapt to the proposed temporal decomposition framework. Refer to subsection C for details.

ii) Since the integration of the strong temporal coupling energy resources, such as ESSs and thermal loads, the convergence performance of the algorithm in [16] is deteriorated. It is necessary to enhance the convergence performance utilizing some acceleration strategies, which are presented in subsection D.

Due to space limitation, we only give the algorithm procedure of the multi-parametric programming projection decomposition algorithm here, as presented in Fig. 2. See *Appendix A* in the supplementary document [20] for detailed algorithm description.

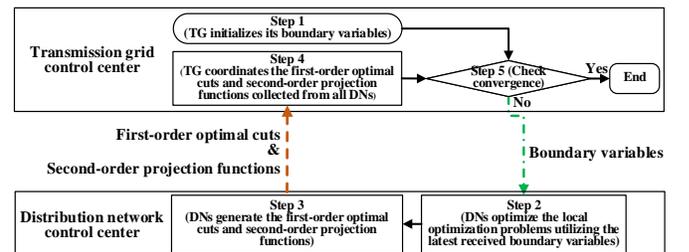

Fig. 2. The iterative procedure of the multi-parametric programming projection based spatial decomposition algorithm.

For the convenience of subsequent descriptions, some key formulas are given here. In Step 3, the expressions of first-order optimal cut $JF^{\text{dist},k}\left(l^{\text{trans},k}\right)^{\left(f^{\text{dist},k,(m)}\right)}$ and second-order projection function $JS^{\text{dist},k}\left(l^{\text{trans},k}\right)^{(m)}$ generated by the $k^{\text{th}}$ DN, are described in formulas (39) and (40), respectively.

$$JF^{\text{dist},k}\left(l^{\text{trans},k}\right)^{\left(f^{\text{dist},k,(m)}\right)} = C^{\text{dist},k}\left(x^{\text{dist},k,(m)}\right) \\ -\left(\lambda_{\text{Bou}}^{\text{dist},k,(m)}\right)^T \left(l^{\text{trans},k}-l^{\text{trans},k,(m)}\right) \quad (39)$$

$$JS^{\text{dist},k}\left(l^{\text{trans},k}\right)^{(m)} = C^{\text{dist},k}\left(x^{\text{dist},k,(m)}\right)-\left(\lambda_{\text{Bou}}^{\text{dist},k,(m)}\right)^T\left(l^{\text{trans},k}-l^{\text{trans},k,(m)}\right) \\ + \frac{1}{2}\left(l^{\text{trans},k}-l^{\text{trans},k,(m)}\right)^T Q\left(l^{\text{trans},k}-l^{\text{trans},k,(m)}\right) \quad (40)$$

where $f^{\text{dist},k,(m)}$ denotes the number of first-order optimal cuts generated until the $m^{\text{th}}$ iteration by the $k^{\text{th}}$ DN. $(\cdot)^{(m)}$ denotes a scalar, vector or matrix calculated in the $m^{\text{th}}$ iteration. The formula of hessian matrix $Q$ is given in subsection C.

In Step 4, the coordinated optimization problem of TG is presented as follows:

$$(\text{P2}) \min_{x^{\text{trans}},l^{\text{trans}},O^{\text{dist},k}}\left(C^{\text{trans}}\left(x^{\text{trans}}\right)+\sum_{k\in DIST}O^{\text{dist},k}\right)$$

$$s.t.\ G^{\text{trans}}\left(x^{\text{trans}},l^{\text{trans}}\right)\leq 0 \quad (41)$$

$$O^{\text{dist},k}\geq JS^{\text{dist},k}\left(l^{\text{trans},k}\right)^{(m)},\forall k\in DIST$$

$$O^{\text{dist},k}\geq JF^{\text{dist},k}\left(l^{\text{trans},k}\right)^{(i)},\forall i=1,2,...,f^{\text{dist},k,(m)}-1,\forall k\in DIST$$

where $O^{\text{dist},k}$ is an intermediate variable representing the objective function of the $k^{\text{th}}$ DN.

### C. Temporal Decomposition of TG's Multi-period ED Model
#### 1) Modeling the Temporal Coupling Constraints Using Overlapping Time Intervals

Suppose that the whole optimization problem of TG is decomposed into $N$ subproblems, denoted as $\{SP_1,…, SP_{n-1}, SP_n, SP_{n+1},…, SP_N\}$. Since the number of subproblems will influence the efficiency of the temporal decomposition algorithm, $N$ should be tuned. Firstly, the time length of every subproblems should be identical to make the subproblems have similar scales and computational complexities. Secondly, a **tuning scheme** is proposed to determine a proper subproblem number $N$, which essentially makes a trade-off between the A-ADMM iteration number and the single solving time of TG's subproblems by analyzing the characteristics of net load profile. The detailed tuning scheme for $N$ is given in the *Appendix B* in the supplementary document [20].

Without loss of generality, we take three consecutive subproblems $SP_{n-1}$, $SP_n$ and $SP_{n+1}$ as an example to make an illustration. To guarantee the optimality of the TG's solution from the perspective of the whole operating horizon, we introduce the **overlapping time interval** to model the temporal coupling constraints between adjacent subproblems. As illustrated by Fig. 3, the first interval of every subproblem, except for the first subproblem, is introduced as an overlapping time interval for two consecutive subproblems. The variables and constraints of every overlapping time interval are duplicated and assigned to the left-side adjacent subproblem to make them independent. Then, the subproblem $SP_n$ can be reformulated as:

$$(\text{P3}) \min_{x_n^{\text{trans}},l_n^{\text{trans}},O_n^{\text{dist},k}}\left(C_n^{\text{trans}}\left(x_n^{\text{trans}}\right)+\sum_{k\in DIST}O_n^{\text{dist},k}\right) \quad (42)$$

$$s.t.\ G_n^{\text{trans}}\left(x_n^{\text{trans}},l_n^{\text{trans}}\right)\leq 0$$

$$O_n^{\text{dist},k}\geq JS_n^{\text{dist},k}\left(l_n^{\text{trans},k}\right)^{(m)},\forall k\in DIST \quad (43)$$

$$O_n^{\text{dist},k}\geq JF_n^{\text{dist},k}\left(l_n^{\text{trans},k}\right)^{(i)},\forall i=1,2,...,f^{\text{dist},k,(m)}-1,\forall k\in DIST$$

where $(\cdot)_n$ represents a scalar, vector or matrix corresponding to subproblem $SP_n$. The expressions of $JS_n^{\text{dist},k}\left(l_n^{\text{trans},k}\right)$ and $JF_n^{\text{dist},k}\left(l_n^{\text{trans},k}\right)$ are detailed in the next subsection.

$$x_n^{\text{trans}} = \{x^{\text{trans}}(t)\}, l_n^{\text{trans}} = \{l^{\text{trans}}(t)\}, l_n^{\text{trans},k} = \{l^{\text{trans},k}(t)\}, \\ \forall t=\{\underbrace{t_{n-1}+1}_{\text{overlapping time interval}},t_{n-1}+2,...,t_n,\ \underbrace{t_n+1}_{\text{overlapping time interval}}\} \quad (44)$$

Fig. 3. Decomposing consecutive subproblems with overlapping time intervals.

The temporal coupling constraints between adjacent subproblems include the generators' ramp up and down constraints (13), and the ESS energy constraints (16). For convenience, a coupling variable indicated by subscript $(\cdot,\cdot)$ is introduced as a decision variable determined by the left-side subproblem. For example, $p_{\text{G},i,(n-1,n)}^{\text{trans}}(t_{n-1}+1)$ and $p_{\text{G},i,(n,n-1)}^{\text{trans}}(t_{n-1}+1)$ are the coupling variables handled by $SP_{n-1}$ and $SP_n$, respectively. Then, we have the following two vectors to represent the coupling variables for $SP_n$.

$$\gamma_{(n,n-1)}^{\text{trans}} = \{p_{\text{G},i,(n,n-1)}^{\text{trans}}(t_{n-1}+1), E_{\text{ESS},i,(n,n-1)}^{\text{trans}}(t_{n-1}), E_{\text{ESS},i,(n,n-1)}^{\text{trans}}(t_{n-1}+1), \\ p_{\text{ESS},i,(n,n-1)}^{\text{dc,trans}}(t_{n-1}+1), p_{\text{ESS},i,(n,n-1)}^{\text{ch,trans}}(t_{n-1}+1), p_{\text{DG},i,(n,n-1)}^{\text{trans}}(t_{n-1}+1), p_{\text{B},i,(n,n-1)}^{\text{trans}}(t_{n-1}+1)\} \quad (45)$$

$$\gamma_{(n,n+1)}^{\text{trans}} = \{p_{\text{G},i,(n,n+1)}^{\text{trans}}(t_n+1), E_{\text{ESS},i,(n,n+1)}^{\text{trans}}(t_n), E_{\text{ESS},i,(n,n+1)}^{\text{trans}}(t_n+1), \\ p_{\text{ESS},i,(n,n+1)}^{\text{dc,trans}}(t_n+1), p_{\text{ESS},i,(n,n+1)}^{\text{ch,trans}}(t_n+1), p_{\text{DG},i,(n,n+1)}^{\text{trans}}(t_n+1), p_{\text{B},i,(n,n+1)}^{\text{trans}}(t_n+1)\} \quad (46)$$

Furthermore, the subproblem $SP_n$ can be rewritten as:

(P4)  min (42)
s.t.  (43)-(44)

$$\gamma_{(n-1,n)}^{\text{trans}} = \gamma_{(n,n-1)}^{\text{trans}} \quad (47)$$

$$\gamma_{(n,n+1)}^{\text{trans}} = \gamma_{(n+1,n)}^{\text{trans}} \quad (48)$$

#### 2) Temporally Decomposed First-order Optimal Cuts and Second-order Projection Functions

The subproblem $SP_n$ is taken as an example to illustrate the temporal decomposition process. Combining (39) and (40), we have the following formulas for $SP_n$.

$$JF_n^{\text{dist},k}\left(l_n^{\text{trans},k}\right)^{\left(f^{\text{dist},k,(m)}\right)} = C_n^{\text{dist},k}\left(x_n^{\text{dist},k,(m)}\right) \\ -\left(\lambda_{\text{Bou},n}^{\text{dist},k,(m)}\right)^T\left(l_n^{\text{trans},k}-l_n^{\text{trans},k,(m)}\right) \quad (49)$$

$$JS_n^{\text{dist},k}\left(l_n^{\text{trans},k}\right) = C_n^{\text{dist},k}\left(x_n^{\text{dist},k,(m)}\right)-\left(\lambda_{\text{Bou},n}^{\text{dist},k,(m)}\right)^T\left(l_n^{\text{trans},k}-l_n^{\text{trans},k,(m)}\right) \\ + \frac{1}{2}\left(l_n^{\text{trans},k}-l_n^{\text{trans},k,(m)}\right)^T Q_n\left(l_n^{\text{trans},k}-l_n^{\text{trans},k,(m)}\right) \quad (50)$$

where $Q_n$ is constructed by the corresponding part of the matrix $Q$ related to subproblem $SP_n$, that is, the diagonal block bounded by the $(t_{n-1}+1)^{th}$ to $(t_n+1)^{th}$ rows and the $(t_{n-1}+1)^{th}$ to $(t_n+1)^{th}$ columns of $Q$, as presented in Fig. 4. And $Q$ can be expressed as:

$$Q = \begin{bmatrix} \dfrac{dx^{dist,k}}{d(l^{trans,k})} \\ I \end{bmatrix}^T \begin{bmatrix} \dfrac{\partial^2 L}{\partial (x^{dist,k})^2} & \\ & \dfrac{\partial^2 L}{\partial (l^{trans,k})^2} \end{bmatrix} \begin{bmatrix} \dfrac{dx^{dist,k}}{d(l^{trans,k})} \\ I \end{bmatrix} \quad (51)$$

where $I$ denotes the identity matrix, whose dimension is $T_{hor}$. The expression of the first-order derivative $\dfrac{dx^{dist,k}}{d(l^{trans,k})}$ is given in the Appendix A in the supplementary document [20]. The Lagrangian function $L$ is given as follows:

$$L(x^{dist,k}, \lambda^{dist,k}) = C^{dist,k}(x^{dist,k}) + (\lambda^{dist,k})^T G^{dist,k}(x^{dist,k}, l^{trans,k,(m)}) \quad (52)$$

i) For the first-order optimal cuts (49), $\lambda_{Bou,n}^{dist,k,(m)}$ is a vector, which can be easily decomposed in temporal dimension. The objective function $C^{dist,k}(\cdot)$ is also decomposable, as it is obtained by accumulating over the whole optimization horizon.

ii) When it comes to the second-order projection function (50), considering that the constraints of DNs' multi-period ED problem are all linear, thus we have:

$$\dfrac{\partial^2 L}{\partial (l^{trans,k})^2} = 0 \quad (53)$$

Moreover, we know that except for variable $p_{DG,i}^{dist,k}(t)$, which has the corresponding quadratic term in the objective function, the second-order derivatives $\dfrac{\partial^2 L}{\partial (x^{dist,k})^2}$ of the other variables of the $k^{th}$ DN are 0. Therefore, we can obtain the following formula.

$$\dfrac{\partial^2 L}{\partial (x^{dist,k})^2} = \begin{bmatrix} 2 \cdot \sigma_{DG} \cdot diag\left(\dfrac{1}{\overline{P}_{DG,i}^{dist,k}}\right) & \\ & 0 \end{bmatrix} \quad (54)$$

where $x^{dist,k} = \begin{bmatrix} p_{DG,i}^{dist,k} & x_{rest}^{dist,k} \end{bmatrix}$. $p_{DG,i}^{dist,k} = \{p_{DG,i}^{dist,k}(t)\}, \forall t \in T$. $x_{rest}^{dist,k}$ refers to all other variables except for $p_{DG,i}^{dist,k}$ of the $k^{th}$ DN. And $diag\left(\dfrac{1}{\overline{P}_{DG,i}^{dist,k}}\right)$ denotes the diagonal matrix consisted of $\dfrac{1}{\overline{P}_{DG,i}^{dist,k}}$.

Combining equation (53) and (54), the expression of $Q$ can be rewritten as:

$$Q = 2 \cdot \sigma_{DG} \cdot \begin{bmatrix} \dfrac{d(p_{DG,i}^{dist,k})}{d(l^{trans,k})} \end{bmatrix}^T \cdot diag\left(\dfrac{1}{\overline{P}_{DG,i}^{dist,k}}\right) \cdot \begin{bmatrix} \dfrac{d(p_{DG,i}^{dist,k})}{d(l^{trans,k})} \end{bmatrix} \quad (55)$$

From (55), we can see that $Q$ is only changed by $\dfrac{d(p_{DG,i}^{dist,k})}{d(l^{trans,k})}$, since $\sigma_{DG}$ and $\dfrac{1}{\overline{P}_{DG,i}^{dist,k}}$ are given. As for the active power of temporal coupling energy resources, i.e. ESSs and thermal loads, they affect $p_{DG,i}^{dist,k}$ through the power flow constraint (22), which thus has an indirect influence on $Q$. Accordingly, we know that the non-diagonal elements in matrix $Q$ correspond to the temporal coupling constraints, that is, formulas (27) and (32).

Then we take subproblem $SP_n$ of TG as an example for illustration. From (50), the elements corresponding to subproblem $SP_n$ are contained in the $(t_{n-1}+1)^{th}$ to $(t_n+1)^{th}$ rows and the $(t_{n-1}+1)^{th}$ to $(t_n+1)^{th}$ columns of $Q$. From (27) and (32), we can know that the decision variables of ESSs and thermal loads during period $t$ are influenced by that of the preceding periods. And the farther away from period $t$, the little the influence tends to be. Based on this, it can be indicated that the elements of the slash-filled parts in Fig. 4 are much smaller than those in the diagonal blocks, because the elements of the slash-filled parts in $Q$ are farther away from the diagonal elements compared with the ones in the diagonal blocks. Therefore, the elements of the slash-filled parts can be mitted to make $Q$ separatable while most information is remained. The case studies also indicate the good convergence performance of the multi-parametric programming projection decomposition algorithm after implementing the temporal decomposition.

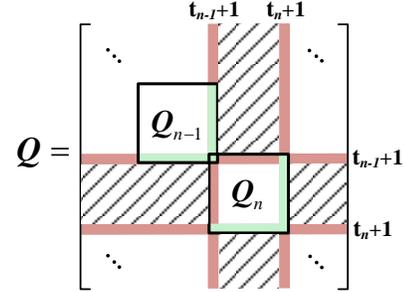

Fig. 4. Illustration of the relative relationships between matrices $Q$, $Q_{n-1}$ and $Q_n$.

*3) The A-ADMM Based Temporal Decomposition Algorithm*

The A-ADMM algorithm [21] can improve the convergence speed significantly. This algorithm introduces the history information into the next iteration by adopting the Nesterov accelerated gradient descent [15]. Due to the limited space, the detailed description of the A-ADMM based temporal decomposition algorithm is given in the *Appendix C* in the supplementary document [20].

*D. The Estimation Method for Determining the Feasible Range of Boundary Variables Utilizing the Aggregate Equivalence Approach*

It is worthy to note that most distributed algorithms are sensitive to initial values. To enhance the convergence performance of the proposed spatio-temporal decomposition method further, the ***aggregate equivalence approach*** is used for determining the feasible range of boundary variables between TG and DNs by analyzing the characteristics of DNs' constraints. This strategy is executed in the initialization stage, then the obtained bounded feasible range of the boundary variables can be added into the C-ED problem , which plays a role in the whole iterative process. Moreover, the case studies verify its significantly effectiveness in enhancing the convergence performance. The process of the aggregate equivalence approach is shown in Fig. 5.

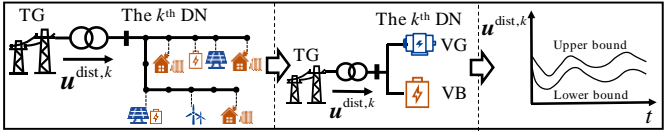

Fig. 5. Illustration of the estimation method for determining the feasible range of boundary variables utilizing the aggregate equivalence approach.

Due to space limitation, here we only introduce the main ideas, please refer to our previous work [22] for details.

i) Firstly, the DERs of every DN are classified into two categories. One category contains the generator-like DERs, including renewable DGs and uncontrollable loads. These DERs have upper and lower limits on the output active power, and some of them have ramping limits. The other category includes the battery-like DERs, including ESSs and thermal loads, which have temporal coupling constraints. Then, these two types of DERs are aggregated into a virtual generator and virtual battery, respectively.

ii) Subsequently, the feasible range of boundary active power $u^{\text{dist},k}$ of the $k^{\text{th}}$ DN can be obtained by summing the flexibility of these two virtual devices.

Then, the bounded feasible range for the boundary transferred active power of every DN is sent to TG. In this way, the iterative process of the proposed algorithm can be accelerated, through adding the bounded feasible range of the boundary variables into the C-ED problem, as shown by (21).

*E. The Spatio-temporal Decomposition Algorithm for the C-ED Problem*

Combining the aforementioned spatial decomposition algorithm, temporal decomposition algorithm, and the aggregate equivalence approach, the spatio-temporal decomposition algorithm is developed for the C-ED problem, whose detailed procedure is given as follows:

**Algorithm 1** Pseudocode for the proposed spatio-temporal decomposition algorithm for the C-ED problem

1. Every DN estimates the feasible range of its boundary variables utilizing the aggregate equivalence approach, and sends the obtained feasible range to TG.
2. TG determines the value of sub-horizon number $N$ by comprehensively considering the net load curve and the calculation burden of every subproblem.
3. Initialize the iteration number $m = 0$.
4. **Loop**
5. Increase the iteration number $m$ by 1, i.e., $m = m + 1$.
6. **While** the convergence criterion of A-ADMM is not met, **do**
7.     TG solves its ED subproblems in parallel utilizing the A-ADMM, which is presented in the *Appendix C* in the supplementary document [20].
8. **end while**
9. TG concatenates the boundary power $l_n^{\text{trans.}(m)}, \forall n \in [1, N]$, forming $l^{\text{trans.}(m)}$, and sends the latest corresponding boundary variable values to DNs in parallel. *// communication*
10. Every DN verifies whether the newly received boundary variable vector is same as that in the last iteration. Take the $k^{\text{th}}$ DN as an example, if the two are same, then the $k^{\text{th}}$ DN skips **Step 11**; If not, the $k^{\text{th}}$ DN optimizes its local problem utilizing the newly received boundary variable values.
11. Every DN computes its first-order optimal cut (39) and second-order projection function (40), and transmits them to TG. *// communication*
12. TG decomposes the received first-order optimal cuts and second-order projection functions into the corresponding subproblems.
13. Convergence condition of the spatial decomposition algorithm is checked by TG which is detailed in the *Appendix A* in the supplementary document [20]. If the condition is satisfied, **output the optimal solution and go to the end**; otherwise, the procedure returns to the **Step 5**.
14. **end loop**

*F. Convergence and Optimality*

Combining the properties of the multi-parametric programming projection decomposition algorithm [16] and the ADMM algorithm [21], we have the following proposition.

***Proposition 1***: For bilevel convex optimization problems with continuously differentiable objectives and constraints, the proposed spatio-temporal decomposition method is guaranteed to converge to the convergence criterion within finite iterations, and the converged solution is globally optimal.

**Proof**: In the temporal dimension, the convergence proof of the A-ADMM algorithm is detailed in [21]. Then, we can conclude that the optimal solution of TG obtained after implementing the temporal decomposition is guaranteed to be consistent with that of TG's original optimization problem, which is ensured by the property of A-ADMM.

In the spatial dimension, our previous study [16] has demonstrate that the convergence and optimality of the proposed multi-parametric programming projection decomposition algorithm is ensured by the first-order optimal cuts for convex optimization problems. As for the second-order projection functions, they are utilized to improve the convergence performance for their better approximation accuracy of DNs' objectives. Meanwhile, it should be noted that conducting the temporal decomposition for the first-order optimal cuts is information lossless, as discussed in the subsection C. Therefore, we can arrive at the conclusion that the convergence of the spatio-temporal decomposition is ensured.

In totally, the convergence and optimality of the proposed spatio-temporal decomposition is theoretically guaranteed. ∎

## IV. NUMERICAL TESTS

*A. Numerical Simulation Settings*

We give the numerical results of the proposed spatio-temporal decomposition method (hereafter named ***proposed method***) with three power systems of different scales.

i) System #1 includes 26 buses in sum, which contains a 14-bus TG and three 4-bus DNs. 4 renewable DGs and 3 ESSs are configured in TG. Moreover, 2 renewable DGs, accompanied with 2 ESSs and 2 residential buildings are supplemented into every DN, in which the household number varies from 10 to 20.

ii) System #2 has 564 buses, which contains a 300-bus TG and eight 33-bus DNs. 14 renewable DGs and 6 ESSs are equipped in TG. Besides, 6 renewable DGs, 3 ESSs and 8 residential buildings are configured in every DN.

iii) System #3 is much larger and contains 3892 buses in total, which includes a 1354-bus TG and eighteen 141-bus DNs. 45 renewable DGs and 13 ESSs are supplemented into TG. In addition, 23 renewable DGs, 7 ESSs and 14 residential buildings are equipped in every DN.

For system #1 and system #2, tolerance $\varepsilon$ is set as $10^{-2}$. As for system #3, tolerance $\varepsilon = 10^{-1}$ is a practical choice. And $c_{\text{pen}}^{\text{dist},k} = 10^5$, $T_{\text{hor}} = 96$, $\Delta T = 15\,\text{min}$, $\sigma_{\text{DG}} = 10^2$, $\sigma_{\text{ESS}} = 10^{-2}$. Detailed data of the cases and parameters of the renewable DGs, ESSs and thermal loads are given in [23]. Case studies are implemented on a laptop with an Intel i7-10875H CPU and 24GB RAM. The C-ED problem is solved by Gurobi [24].

## B. Performance of the Proposed Spatio-temporal Decomposition Method

### 1) Exactness of the Proposed Method Compared With the Centralized Method

The optimal objective function values of our proposed method are compared to the centralized method to verify whether the proposed method could obtain the optimal solution for C-ED problem. As illustrated in Table I, the almost identical results demonstrate the proposed method's exactness and validity. Moreover, Table II presents the root-mean-square errors on the boundary active power of the proposed method compared to the centralized method. We can observe that the errors are considerably small.

TABLE I
COMPARISONS OF THE OBJECTIVE FUNCTION VALUES BETWEEN THE PROPOSED METHOD AND THE CENTRALIZED METHOD

| Cases | Proposed method | Centralized method |
|---|---|---|
| System #1 (26 buses) | $207986.8 | $207982.9 |
| System #2 (564 buses) | $60865272.1 | $60865211.5 |
| System #3 (3892 buses) | $149545030.9 | $149544902.2 |

TABLE II
ROOT-MEAN-SQUARE ERRORS ON THE BOUNDARY ACTIVE POWER OF THE PROPOSED METHOD COMPARED TO THAT OF THE CENTRALIZED METHOD

| Cases | Root-mean-square errors |
|---|---|
| System #1 (26 buses) | $1.13 \times 10^{-2}$ |
| System #2 (564 buses) | $2.47 \times 10^{-2}$ |
| System #3 (3892 buses) | $5.39 \times 10^{-2}$ |

### 2) Comparing the Proposed Method With Other Methods on the Coordination Performance

The coordination performance of the proposed method is compared with that of the multi-parametric programming projection decomposition algorithm (hereafter named ***multi-parametric method***) proposed in our previous research [16] and the GBD algorithm [1] in terms of iteration numbers and computational time. Specifically, the total computational time refers to the time taken to converge to the given tolerance for different methods running in parallel. The comparison results are presented in Table III.

As listed in Table III, the proposed method is significantly superior to the GBD algorithm in terms of the iteration numbers and computational efficiency. Obviously, the larger the scale of power system is, the greater the advantage of the proposed method is. Besides, we can also see that the proposed method has better coordination performance compared with the multi-parametric method. There are three reasons why the proposed method exhibits considerable coordination advantages:

i) The proposed method introduces the second-order projection functions, which can enhance the coordination efficiency considerably, contrast to the GBD with only first-order convergence.
ii) Through decomposing the whole scheduling horizon into several sub-horizons, the computational efficiency of TG is further improved.
iii) By introducing the bounded feasible region of boundary variables estimated by the aggregate equivalence approach, the proposed method can quickly step out of the infeasible region of the C-ED problem, and speed up the iterative process.

TABLE III
PERFORMANCE COMPARISONS BETWEEN THE PROPOSED METHOD AND OTHER METHODS

| Cases | Performance | Proposed method | Multi-parametric method | GBD |
|---|---|---|---|---|
| System #1 (26 buses) | Iteration numbers* | 6 | 15 | 33 |
| | Calculation time (s) | 0.36 | 0.92 | 2.04 |
| System #2 (564 buses) | Iteration numbers* | 9 | 34 | 52 |
| | Calculation time (s) | 3.93 | 37.4 | 57.2 |
| System #3 (3892 buses) | Iteration numbers* | 19 | 48 | 94 |
| | Calculation time (s) | 38.96 | 1802.6 | 3530.1 |

* The iteration numbers correspond to that of the spatial decomposition algorithm.

### C. Performance Comparisons of the Proposed Method With and Without Temporal Decomposition

We compare the performance of the proposed method with and without temporal decomposition, as illustrated in Fig. 6. It shows that the proposed temporal decomposition can dramatically reduce the computation time especially for large scale systems. The reasons are twofold:

i) The proposed temporal decomposition method makes the optimization problem of TG can be implemented in a parallel manner efficiently.
ii) The A-ADMM algorithm has a good convergence speed. The reason is that it not only employs the Nesterov accelerated gradient descent method, but also adopts some acceleration strategies. Specifically, the warm-start strategy is developed to obtain a good starting point, and the adaptive updating of penalty parameter [25] is introduced to decrease the influence of hyperparameter selection on the algorithm convergence.

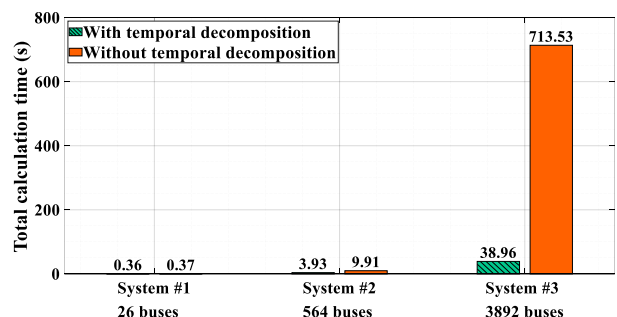

Fig. 6. Comparison of the performances of the proposed method with and without temporal decomposition.

### D. Demonstrating the Performance of the Presented Aggregate Equivalence Approach

We compare the computational performance of the proposed spatial-temporal decomposition algorithm with and without the aggregate equivalence approach. Fig. 7 illustrates that the introduced aggregate equivalence approach can greatly improve the computational efficiency. This is because the proposed method can quickly jump out of the infeasible region of the C-ED problem by utilizing the aggregate equivalence approach. Correspondingly, the time for searching for a feasible solution for the C-ED problem in the initial stage is significantly decreased. Moreover, the iterative process of the proposed algorithm is accelerated, through adding the bounded feasible range of the boundary variables obtained by the aggregate equivalence approach into the C-ED problem.

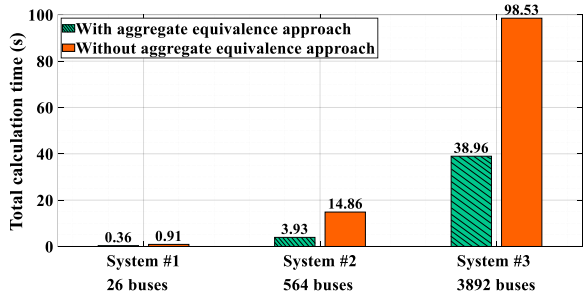

Fig. 7. Comparison of the performances of the proposed method with and without the aggregate equivalence approach.

V. CONCLUSION

With numerous DERs integrated into DNs, the C-ED problem for the integrated transmission and distribution grids, plays a vital role in the power systems operation. This paper proposes a novel spatio-temporal decomposition method to make the C-ED applicable for large-scale transmission and distribution grids. In the temporal dimension, the multi-period ED of TG is decomposed into several subproblems through introducing auxiliary variables and overlapping time intervals to cope with the temporal coupling constraints. Besides, an A-ADMM based temporal decomposition algorithm with the warm-start strategy is developed to parallelly coordinate the subproblems of TG. In the spatial dimension, a multi-parametric programming projection based spatial decomposition algorithm is proposed to coordinate the ED problems of TG and DNs in a distributed manner, under the proposed temporal decomposition framework. Based on the aggregate equivalence approach, a method estimating the feasible range of boundary variables between TG and DNs is introduced to enhance the spatial decomposition algorithm's convergence performance further. Numerical tests justify the proposed spatio-temporal decomposition method has the potential for solving the real-world problems.